\documentclass[
aps,%
12pt,%
final,%
notitlepage,%
oneside,%
onecolumn,%
nobibnotes,%
nofootinbib,
superscriptaddress,%
noshowpacs,%
centertags]%
{revtex4}
\RequirePackage{graphicx}

\newcommand{\Phib}{\bar \Phi}

\def\refitem#1{\relax}

\begin{document}
\title{Fluctuations and the QCD phase diagram}

\author{\firstname{Bernd-Jochen} \surname{Schaefer}}
\email{bernd-jochen.schaefer@uni-graz.at}
\affiliation{Institut f\"ur Physik, Karl-Franzens-Universit\"at Graz, 
8010 Graz, Austria}

\begin{abstract}
  In this contribution the role of quantum fluctuations for the QCD
  phase diagram is discussed. This concerns in particular the
  importance of the matter back-reaction to the gluonic sector. The
  impact of these fluctuations on the location of the
  confinement/deconfinement and the chiral transition lines as well as
  their interrelation are investigated. Consequences of our findings
  for  the size of a possible quarkyonic phase and location of a
  critical endpoint in the phase diagram are drawn.
\end{abstract}

\maketitle

\section{Introduction}

Driven by running high-energy heavy-ion collision experiments at LHC,
RHIC and the SPS and upcoming heavy-ion facilities such as FAIR, NICA
and J-PARC there is a strong interest to gain a deeper understanding
of the properties of QCD matter under extreme conditions. The general
chiral and deconfining phase structure and in particular the possible
existence of a critical endpoint (CEP) in the QCD phase diagram are
much under debate, see e.g.~\cite{BraunMunzinger:2009zz}. Analyzing
the relevant temperature and density region in the phase diagram is
clearly a non-perturbative task and suitable methods have to be
applied, e.g.~\cite{Pawlowski:2010ht}.

For vanishing chemical potential, lattice QCD is a suitable
non-perturbative technique to provide reliable insights in the QCD
finite temperature transition but fails at larger densities due to the
notorious fermion sign problem at non-vanishing chemical potential.
Recently, impressive progress has been achieved in the framework of
QCD-motivated Polyakov-loop extended chiral models such as the
Polyakov-Nambu--Jona-Lasinio (PNJL) \cite{Fukushima:2003fw} and the
Polyakov-quark-meson (PQM) \cite{pqm, Schaefer:2007pw, Herbst:2010rf}
models. In these models information about the confining glue sector of
QCD is included in form of an effective Polyakov-loop potential in the
spirit of a Landau-Ginzburg approach together with a coupling of the
Polyakov-loop to the quarks. The parameters of the Polyakov-loop
potential are extracted from pure Yang-Mills lattice simulations at
vanishing densities.

Such models have often been investigated in mean-field approximations
but the matter sector of these models has also been studied in detail
beyond the mean-field level by taking into account the quark-meson
quantum fluctuations mostly in a functional renormalization group
(FRG) approach, see e.g. \cite{Schaefer:2004en}.

In this context, the quantum back-reaction of the matter sector to the
gluonic sector in these Polyakov-loop extended models has been
omitted. The Polyakov-loop potential has a parameter $T_0$ that is
adjusted to the, for $N_c = 3$, first-order deconfinement transition
temperature of $T_c = 270$ MeV in the pure gauge theory where no
dynamical quarks are present. In \cite{Schaefer:2007pw} 
the quantum back-reaction has been taken into account and acquires in
the presence of dynamical quarks a quark flavor and quark chemical
potential dependent transition temperature, $T_0 (N_f,\mu)$, in the
Polyakov-loop potential. This first solid phenomenological estimate of
the $N_f$- and $\mu$ dependence of $T_0$ is based on HTL/HDL
perturbation theory and one-loop running of the QCD $\beta$-function.
It is remarkable, that this first estimate has been confirmed by first
principle QCD computations with the FRG at real and imaginary chemical
potential \cite{Braun:2009gm} and also in a combination of the PNJL
with the statistical model \cite{Fukushima:2010is}.

This already nicely demonstrates that valuable and model-parameter
independent information on the QCD phase diagram can be obtained with
these QCD-motivated effective models and that these models can be
understood as specific controlled approximations to full dynamical
QCD. Furthermore, this allows to systematically advance these models
towards full dynamical QCD.

\section{Beyond mean-field}

The method of choice for including quantum fluctuations at
non-vanishing temperature and density is the functional
renormalization group (FRG) technique.

\begin{figure}[h]
    \centering
        \includegraphics[width=0.9\textwidth]{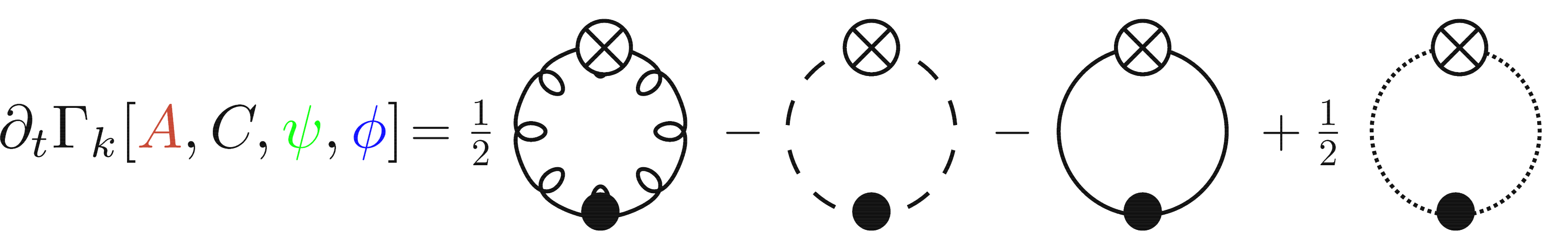}
        \caption{Various (Gluons, ghosts, quarks and mesons) flow
          contributions to the QCD FRG for the effective action: The
          loops with the black bullet denote the corresponding full
          field dependent propagators and the crosses the cutoff
          regulator insertions. See text for details.}
  \label{fig:QCDflow}
\end{figure}

However, in a full QCD computation the resulting FRG flow equation for
the effective action consists of several flow contributions and is
provided by the diagrammatic form as shown in Fig.~\ref{fig:QCDflow},
see \cite{Braun:2009gm}. The first two loops depict the flow
contributions of the full gluon ($A$) and ghost ($C$) propagators with
a regulator denoted as a cross in the loops. These two flow diagrams
represent the quantum fluctuations of the pure glue sector of QCD and
generate the Polyakov-loop potential. The third loop ($\psi$) stands
for the quark fluctuations and the last contribution ($\phi$) encodes
the mesonic fluctuations which are generated by dynamical
hadronization, see e.g.~\cite{Gies:2001nw} for details.

\begin{figure}[h]
    \centering
        \includegraphics[width=0.25\textwidth]{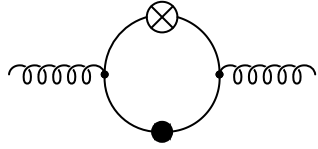}
        \caption{Quark contributions to the flow of the gluon
          propagator.}
  \label{fig:quarkflow}
\end{figure}

We emphasize that the flow equation for the effective action is a
highly non-linear coupled system. An example of this fact is given in
Fig.~\ref{fig:quarkflow} where the gluonic (and indirectly also the
ghost) propagator is modified in the presence of dynamical quarks. In
addition, there are also gluonic contributions to the quark and meson
correlation functions. Since the first two loops generate the pure
Yang-Mills Polyakov-loop potential these observations elucidate the
flavor and density dependence of the $T_0$ parameter in the
Polyakov-loop potential \cite{Pawlowski:2010ht, Schaefer:2007pw}. On
the other hand, if the first two diagrams, i.e., the gluon and ghost
loops, are neglected in the full flow equation, the quantum dynamics
of the quark-meson system is obtained. For recent works in the FRG
framework see, e.g., \cite{Schaefer:2004en}. This simple additive
structure of the different contributions to the full QCD flow shows
the inherent power of quark-meson-type models to be systematically
improved towards full QCD within a FRG picture. Conversely, it also
allows us to use directly full QCD information in connection with
these effective models which further reduce their parameter dependency
and enhance their predictive power. An example is the inclusion of the
$N_f$ and $\mu$-dependence in $T_0$ in the Polyakov-loop potential
together with the dynamics of the quark-meson sector of the PQM model
in the presence of a non-trivial Polyakov-loop expectation value
\cite{Herbst:2010rf}.

\section{Phase structure}

\begin{figure}[h]
    \centering
        \includegraphics[width=0.45\textwidth]{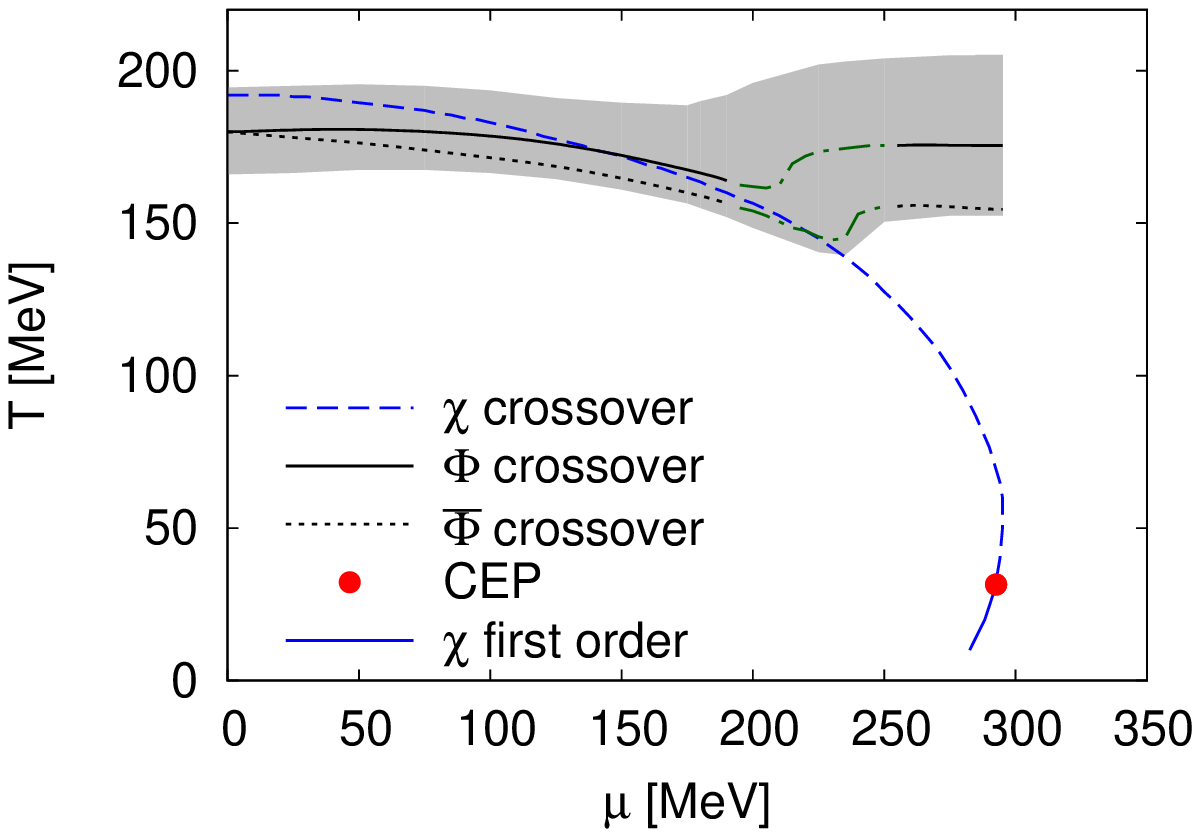}
        \includegraphics[width=0.45\textwidth]{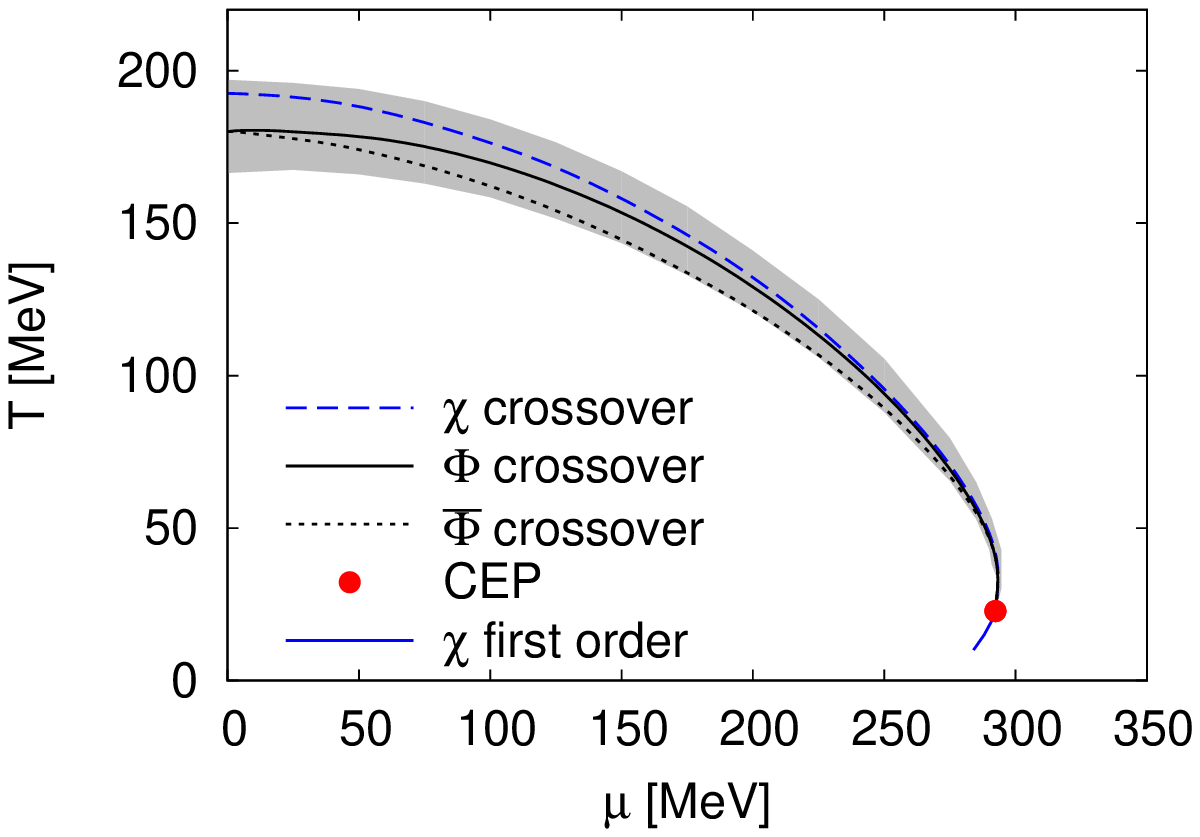}
        \caption{The chiral ($\chi$) and deconfinement ($\Phi$,
          $\Phib$) phase diagrams for a constant $T_0 = 208$ MeV (left
          panel) and for a density dependent $T_0(\mu)$ (right panel).
          The grey band corresponds to the width of the $d\Phi/dT$ at
          80\% of its peak height. In the left panel, the kink
          behaviour in the $\Phi$, $\Phib$ lines around
          $\mu \sim 200 - 250$ MeV  vanishes if the used approximation of
          the effective potential on the mean-field solutions of the
          Polyakov-loop EoMs is improved.  }
\label{fig:phasediagram}
\end{figure}

The resulting phase diagrams in a PQM truncation without (left panel)
and with (right panel) the back-reaction of the matter sector to the
glue sector, encoded in $T_0 (N_f, \mu)$, are displayed in
Fig.~\ref{fig:phasediagram}. The right panel shows the phase diagram
with the full dynamics in the sense that the quark-meson dynamics in
the presence of the Polyakov-loop background is included with the FRG
together with the matter back-reaction. Comparing both phase diagrams,
the grey band which corresponds to the width of the temperature
derivative of the Polyakov-loop variable at 80\% of its peak height,
shrinks with increasing $\mu$ if the back-reaction is included. As a
consequence, the deconfinement transition gets sharper at larger $\mu$
and stays close to the chiral phase boundary. This difference is even
more pronounced if the quantum fluctuations are neglected via a
mean-field PQM calculation with a constant $T_0$
\cite{Schaefer:2009ui}. For physical quark masses the chiral
(pseudo)critical transition lines curve downwards with increasing
chemical potential but there is a large region in which chiral
symmetry is restored and matter confined (in the statistical sense)
\cite{McLerran:2007qj}. Thus, the size of this region depends strongly
on the Polyakov-loop dynamics: it has been shown in
\cite{Schaefer:2009ui} within a mean-field PQM calculation with the
matter back-reaction, i.e., with a running $T_0(\mu)$, the size of
this region has considerably shrunk.

Fig.~\ref{fig:phasediagram} highlights the importance of the matter
back-reaction via a running $T_0(\mu)$: with a constant $T_0$ (left
panel) the Polyakov loops $\Phi$ and $\Phib$ start to deviate for
non-vanishing chemical potential and the widths of their temperature
derivatives increase over the whole phase diagram. On the contrary in
a full FRG solution where the quark-meson fluctuations and the matter
back-reaction via a running $T_0(\mu)$ are taken into account both
transition lines coincide over the whole phase diagram (right panel).

In summary, we found that within the fully coupled system, in
particular, including the back-reaction of the matter fluctuations to
the Yang-Mills sector, the peak locations of the order parameters for
the chiral and deconfinement transitions are closely related.
Of course, at large densities the present truncations
without baryons becomes questionable in particular in the hadronic
phase for smaller temperatures. Nevertheless, two important generic
findings of these FRG analyses can be compiled: the location of a
possible critical endpoint can be excluded for small chemical
potentials and there is little room for a chirally symmetric and
confining region in the phase diagram.

\begin{acknowledgments}
  The work presented here was done in collaboration with Tina K.
  Herbst and Jan M. Pawlowski. The author is grateful to the
  organizers of the CPOD2010 and DM2010 meetings for the invitation
  and hospitality extended to him during his visits at the JINR in
  Dubna, Russia.
\end{acknowledgments}

\newpage

\newpage

\begin{center}
FIGURE CAPTIONS
\end{center}
\begin{enumerate}
\item Various (Gluons, ghosts, quarks and mesons) flow contributions to
  the QCD FRG for the effective action: The loops with the black
  bullet denote the corresponding full field dependent propagators and
  the crosses the cutoff regulator insertions. See text for details.
\item Quark contributions to the flow of the gluon propagator.
\item The chiral ($\chi$) and deconfinement ($\Phi$,
          $\Phib$) phase diagrams for a constant $T_0 = 208$ MeV (left
          panel) and for a density dependent $T_0(\mu)$ (right panel).
          The grey band corresponds to the width of the $d\Phi/dT$ at
          80\% of its peak height. In the left panel, the kink
          behaviour in the $\Phi$, $\Phib$ lines around
          $\mu \sim 200 - 250$ MeV  vanishes if the used approximation of
          the effective potential on the mean-field solutions of the
          Polyakov-loop EoMs is improved.
          \end{enumerate}

\end{document}